\documentstyle[12pt]{article}
\renewcommand
\baselinestretch{1.5}
\newcommand{\ls}{\lefteqn{<}\raise-4pt\hbox{$\sim$}}
\begin{document}
\begin{flushright}
LAEFF 96/25\\
KSPU-96-03\\
{\tt October 23, 1996}
\end{flushright}
\vspace{.5cm}
\begin{center}
{\LARGE Self-consistent Wormhole Solutions of Semiclassical Gravity }
\end{center}
\vspace{1cm}
\begin{center}
{\large David Hochberg{$^{ a,}$}\footnote{Electronic address:
hochberg@laeff.esa.es}},
{\large Arkadiy Popov{$^{ b,}$}\footnote{Electronic address:
popov@kspu.ksu.ras.ru}
 and Sergey V. Sushkov{$^{ b,}$}\footnote{Electronic address:
sushkov@kspu.ksu.ras.ru}} \\
{\large \sl {$^a$}Laboratorio de Astrof\'{\i}sica Espacial
y F\'{\i}sica Fundamental\\
Apartado 50727, 28080  Madrid, Spain}\\
{\large \sl {$^b$}Department of Mathematics, Kazan State Pedagogical
University\\ Mezhlauk st., 1, Kazan 420021, Russia}
\end{center}
\begin{abstract}
We present the first results of a self-consistent solution of
the semiclassical
Einstein field equations corresponding to
a Lorentzian wormhole coupled to a quantum scalar field. 
The specific solution presented here represents a wormhole connecting
two asymptotically spatially flat regions. 
In general, the diameter of the wormhole
throat, in units of the Planck length, can be arbitrarily large, depending
on the values of the scalar coupling $\xi$ and the boundary values for the
shape and redshift functions.
In all cases we have considered, there is a fine-structure 
in the form
of Planck-scale oscillations or ripples superimposed on the solutions.
\end{abstract}

\newpage
\renewcommand
\baselinestretch{1.5}

Wormholes are topological handles in spacetime linking widely separated regions
of a single universe or ``bridges'' joining two different spacetimes. Interest
in these configurations dates back at least as far as 1916 \cite{Flamm}
with punctuated
revivals of activity following both the classic work of Einstein and Rosen
 in 1935 \cite{Eins} and the later series of works
initiated by Wheeler in 1955 \cite{Wheeler}.
More recently, a fresh interest in the topic has been rekindled by the
work of Morris and Thorne \cite{Mthorne}, leading to a flurry of activity
branching off into diverse directions. A brief resum\'e of current work
devoted to the physics of Minkowski-signature wormholes includes topics
addressing fundamental features of traversable wormholes
\cite{Mthorne,MTY},
explicit
modelling of wormhole metrics
and the corresponding classical \cite{Viss1}
and quantum
mechanical stability \cite{Viss2} analyses, wormholes as time machines and the
problem of causality violation \cite{Viss3},
wormholes in higher-derivative gravity \cite{Hoch1},
wormholes from the gravitationally squeezed vacuum \cite{Hoch2},
possible cosmological
consequences of early universe wormholes \cite{Hoch3,Liu},
and wormholes as gravitational lenses \cite{Cramer}.
A thourough and up-to-date survey of the present status of
Lorentzian wormholes may be found in the excellent monograph by Visser
\cite{Viss4}.

There are plausible physical arguments suggesting that Lorentzian
wormholes should exist at least at scales of order the Planck length.
Most of what is known about them is based on detailed analyses of models, and
within the literature devoted to the subject, the existence of
wormholes is taken as a working hypothesis.
Metrics describing wormholes with desirable traits are
written down by fiat and the properties of the corresponding
hypothetical stress-energy tensors needed to
support the wormhole spacetime are then
worked out and analyzed. 
In an example of an analysis of this sort, Ford and Roman \cite{FordRoman}
have derived approximate constraints on the magnitude and duration
of the negative energy densities which must be observed by a timelike
geodesic observer in static spherically symmetric wormhole spacetimes.
More recently, Taylor, Hiscock and Anderson have argued that stress tensors
for massive minimally and/or conformally coupled scalars fail to meet the
requirements for maintaining five particular types of static spherically
symmetric wormholes, but have not solved the 
backreaction problem \cite{Taylor}.
In particular, no one up to now has succeeded in
writing down a bona-fide wormhole {\em solution} of
either the classical or semiclassical
Einstein field equations. The reason for this state of affairs is easy to
understand. In the first case, it is well known that any stress-energy
that might give rise to a wormhole must violate one or more of the cherished
energy conditions of classical general relativity \cite{Mthorne,MTY}.
Hence wormholes cannot
arise as solutions of classical relativity and matter. If they exist, they must
belong to the realm of semiclassical or perhaps quantum gravity. In
the realm of semiclassical gravity, one sets the Einstein tensor equal to the
expectation value of the
stress-energy tensor operator of the quantized fields present:
\begin{equation}
G_{\mu \nu} = 8\pi \langle T_{\mu \nu} \rangle.
\end{equation}
A primary technical difficulty in semiclassical gravity is that
$\langle T_{\mu \nu} \rangle$ depends strongly on the metric and is generally
difficult to calculate. Until recently, all calculations of
$\langle T_{\mu \nu} \rangle$ have been performed on fixed
classical backgrounds.
The fixed background in turn, as its name implies,
must be a solution of the classical Einstein
equation. As there are no classical wormhole backgrounds, no
corresponding semiclassical back-reaction problem can be set up
meaningfully.

In this Letter we present and summarize
the results of the first {\em self-consistent}
wormhole solutions of semiclassical Einstein gravity.
Prior to this, a self-consistent wormhole solution had been obtained
using a phenomenological stress tensor not derived from quantum field theory
 \cite{Sushkov}.
The results of the present
calculation may be taken as numerical evidence for the existence
of Lorentzian wormholes.
For the source term in (1) we employ the
stress-energy tensor of Anderson, Hiscock and Samuel, which is calculated
for a quantized scalar field in an arbitrary static and spherically
symmetric spacetime \cite{Anderson}.
This means that in the field equation (1), both
the Einstein tensor as well as the stress tensor individually depend
on two independent functions of the radial coordinate.
When supplemented with the appropriate set of boundary
conditions, the solutions of the resultant coupled nonlinear
differential equations
are therefore self-consistent because
both the spacetime metric and the distribution of
stress-energy are determined simultaneously and coherently.
This should be contrasted clearly with, and distinguished from, the approach
taken in perturbative back-reaction problems in which the (background)
spacetime is fixed once and for all,
and the stress tensor is supplied as an explicitly known
function of that fixed background.

In the following, we set up the semiclassical field equations valid for
any static and spherically symmetric spacetime containing
quantized scalar matter
and discuss the nature of
the boundary conditions needed for solving this system of fourth-order
equations. We consider the case of a conformally coupled scalar field.
Results of the numerical calculations are presented graphically and
we appeal to an approximate but
analytic treatment to reveal the important
asymptotic behavior of the solution. Further details of this and related
calculations will appear in a separate paper.

The metric for a general static and spherically symmetric spacetime can be
cast into the form
\begin{equation}
ds^2 = -f(l)dt^2 + dl^2 + r^2(l) \left( d\theta^2 + \sin^2 \theta \, d\phi^2
\right),
\end{equation}
where $f(l),r(l)$ are two independent functions of the proper distance
$l$.
This form of the metric is suitable for dealing with wormholes, or
for that matter, any static and spherically symmetric spacetime which might
contain a throat, i.e., $r(0) > 0$.
Following the established nomenclature, $f$ is denoted
 the redshift function and
$r$ is known as the shape function \cite{Viss4}.
In this metric, the semiclassical field
equations take the form \cite{Units}
\begin{equation}
G^{\nu}_{\mu} (f(l),r(l)) = 8\pi \langle T^{\nu}_{\mu} (f(l),r(l);\xi)
\rangle,
\end{equation}
where $\xi$ is the (non-minimal) scalar coupling to the metric.
Note the dependence of both sides of the equation (3)
on the two unknown functions. In the metric (2), the components of the
Einstein tensor are given by
$G^t_t = \frac{2r''}{r} + \frac{r'^2}{r^2} -\frac{1}{r^2},
G^l_l = \frac{f'r'}{fr} + \frac{r'^2}{r^2} -\frac{1}{r^2},$
and
$G^{\theta}_{\theta} = \frac{f''}{2f} + \frac{r''}{r}
+ \frac{f'r'}{2fr}
 -\frac{f'^2}{4f^2}$;
the prime denotes the derivative with respect to $l$.
An accurate analytic approximation to the exact numerically
calculated scalar field
stress energy tensor was developed in Ref. \cite{Anderson} and is
expressed there as
\begin{equation}
\langle T^{\nu}_{\mu} \rangle_{analytic} = (T^{\nu}_{\mu})_0 +
(\xi - \frac{1}{6})(T^{\nu}_{\mu})_1 +
(\xi - \frac{1}{6})^2(T^{\nu}_{\mu})_2 + (T^{\nu}_{\mu})_{log},
\end{equation}
where the individual factors $(T^{\nu}_{\mu})_{0,1,2}$ are written in terms of
two functions of the radial coordinate.
The last factor $(T^{\nu}_{\mu})_{log}$ was left in terms of
combinations of
curvature tensors and covariant derivatives \cite{Foot1}.
To be of practical use in
the present calculation, we must work out these curvature terms and transform
all the  factors in (4) in terms of our two functions $f(l),r(l)$
of the proper distance $l$.
Carrying out
these straightforward but rather lengthy steps we
find that the semiclassical Einstein equations for a conformally
coupled scalar $(\xi = \frac{1}{6})$
are as follows ($K^2 = \frac{1}{5760 \pi}$):
\vskip .2cm\noindent {\bf the~$tt$~component:}
\begin{eqnarray}
\lefteqn{\frac{2r''}{r} + \frac{r'^2}{r^2} -\frac{1}{r^2}=}&&\\ \nonumber
&&K^2 \Bigg[ \frac{32}{r^4} + 7\frac{f'^4}{f^4}
  - 24\frac{f'^3r'}{f^3r} + 24\frac{f'^2r'^2}{f^2r^2}
  - 32\frac{r'^4}{r^4} +4\frac{f'^2f''}{f^3}
  - 12\frac{f''^2}{f^2} +80\frac{f'^2r''}{f^2r} \\ \nonumber
&&- 160\frac{f'r'r''}{fr^2} +128\frac{r'^2r''}{r^3}
  - 64\frac{f''r''}{fr} + 32\frac{r''^2}{r^2}
  - 16\frac{f'f'''}{f^2} +64\frac{r'f'''}{fr}
  - 96\frac{f'r'''}{fr} -64\frac{r'r'''}{r^2}\\ \nonumber
&&+ 16\frac{f''''}{f} -64\frac{r''''}{r}
  + \ln f \Bigg( \frac{16}{r^4} -49\frac{f'^4}{f^4}
  + 44\frac{f'^3r'}{f^3r} +20\frac{f'^2r'^2}{f^2r^2}
  - 16\frac{r'^4}{r^4} +16\frac{f'^2f''}{f^3}  \\ \nonumber
&&-104\frac{f'r'f''}{f^2r} -36\frac{f''^2}{f^2}
  +8\frac{f'^2r''}{f^2r} -80\frac{f'r'r''}{fr^2}
  +64\frac{r'^2r''}{r^3} + 16\frac{f''r''}{fr}
  +16\frac{r''^2}{r^2} -48\frac{f'f'''}{f^2}\\ \nonumber
&&+64\frac{r'f'''}{fr} -16\frac{f'r'''}{fr}
  - 32\frac{r'r'''}{r^2} + 16\frac{f''''}{f}
  -32\frac{r''''}{r} \Bigg)\Bigg],
\end{eqnarray}
\vskip .2cm\noindent {\bf the~$ll$~component:}
\begin{eqnarray}
\lefteqn{\frac{f'r'}{fr} + \frac{r'^2}{r^2} -\frac{1}{r^2}=}&&\\ \nonumber
&& K^2\Bigg[ \frac{f'^4}{f^4} -16\frac{f'^3 r'}{f^3 r}
+64\frac{f' r'^3}{f r^3} -4\frac{f'^2 f''}{f^3}
 + 64\frac{f'r'f''}{f^2 r} -64\frac{r'^2 f''}{f r^2}
 - 4\frac{f''^2}{f^2}- 48\frac{f'^2 r''}{f^2 r} \\ \nonumber
&& + 32\frac{f'r'r''}{f r^2} + 32\frac{f'' r''}{f r}
   + 8\frac{f' f'''}{f^2} - 32\frac{r'f'''}{fr}
   - 32\frac{f' r'''}{f r} + \ln f \Bigg( \frac{16}{r^4}
   + 7\frac{f'^4}{f^4} -20\frac{f'^3r'}{f^3r} \\ \nonumber
&& - 4\frac{f'^2r'^2}{f^2 r} + 32\frac{f'r'^3}{f r^3}
   - 16\frac{r'^4}{r^4} -12\frac{f'^2f''}{f^3}
   + 48\frac{f'r'f''}{f^2r} -32\frac{r'^2f''}{fr^2}
   - 4\frac{f''^2}{f^2} - 16\frac{f'^2r''}{f^2 r} \\ \nonumber
&& + 16\frac{f'r'r''}{fr^2} + 16\frac{f''r''}{fr}
   - 16\frac{r''^2}{r^2} + 8\frac{f'f'''}{f^2}
   - 16\frac{r'f'''}{fr} - 16\frac{f'r'''}{fr}
   + 32\frac{r'r'''}{r^2} \Bigg) \Bigg],
\end{eqnarray}
\vskip .2cm\noindent {\bf the~$\theta\theta$~component:}
\begin{eqnarray}
\lefteqn{ \frac{f''}{2f} + \frac{r''}{r} + \frac{f'r'}{2fr}
 -\frac{f'^2}{4f^2}=}&&\\ \nonumber
&&K^2\Bigg[ 17\frac{f'^4}{f^4} -16\frac{f'^3r'}{f^3r}
  - 32\frac{f'r'^3}{fr^3}-52\frac{f'^2f''}{f^3}
  + 32\frac{f'r'f''}{f^2r} +32\frac{r'^2f''}{fr^2}
  + 28\frac{f''^2}{f^2}+16\frac{f'^2r''}{f^2r}\\ \nonumber
&&+ 64\frac{f'r'r''}{fr^2} -32\frac{f''r''}{fr}
  + 24\frac{f'f'''}{f^2}-48\frac{r'f'''}{fr}
  + 32\frac{f'r'''}{fr}-16\frac{f''''}{f}
  + \ln f\Bigg( \frac{-16}{r^4} +21\frac{f'^4}{f^4}\\ \nonumber
&&- 12\frac{f'^3r'}{f^3r} -8\frac{f'^2r'^2}{f^2r^2}
  - 16\frac{f'r'^3}{fr^3} +16\frac{r'^4}{r^4}
  - 52\frac{f'^2f''}{f^3} +28\frac{f'r'f''}{f^2r}
  + 16\frac{r'^2f''}{fr^2}+20\frac{f''^2}{f^2}\\ \nonumber
&&+ 4\frac{f'^2r''}{f^2r} +32\frac{f'r'r''}{fr^2}
  - 32\frac{r'^2r''}{r^3}-16\frac{f''r''}{fr}
  + 20\frac{f'f'''}{f^2} -24\frac{r'f'''}{fr}
  + 16\frac{f'r'''}{fr}-8\frac{f''''}{f}
  + 16\frac{r''''}{r} \Bigg)\Bigg].
\end{eqnarray}

The full expressions for the general $\xi$-dependent
components of the Einstein equations are too lengthy to
be shown here.

In order to solve the above field equations, we must supply an appropriate
set of  boundary conditions.
Note that the $tt$ and $\theta \theta$-equations are fourth order differential
equations
in the
two functions $f$ and $r$ while the $ll$-equation is third order. This latter
equation is actually a constraint which plays the role of restricting
the solutions of the coupled fourth order differential equations \cite{Foot2}.
Since we seek wormhole solutions of (1), we
shall specify the boundary data at $l=0$, the origin of proper
distance as measured from the throat.
The complete set of boundary conditions therefore requires specifying
eight pieces of information,
namely  $f(0),f'(0),f''(0),f'''(0)$ as well as
$r(0),r'(0),r''(0)$ and $r'''(0)$.
In particular, $r(0)$ is
the (wormhole) throat
radius, and we can use the $ll$-equation to derive an exact relation
between $r(0)$ and just three initial conditions.
In order to do so, let us recall the
boundary conditions appropriate to a wormhole (or more generally, for any
spacetime with a throat). If a solution of (1) is to possess a throat, then
we must require that $r(0) > 0,\, r'(0) = 0$ and $r''(0) \geq 0$.
In other words, there
must exist a sphere of minimum radius located at the origin of proper
distance. So $r(l)$ is simply a positive increasing function of $l$ in the
neighborhood of the origin.
For two-way passage through the wormhole, it is judicious to
avoid solutions with horizons at the
throat. This is taken care of by requiring a (locally)
nonvanishing redshift function:
$f(0) > 0$ (the other possibility,
$f(0) < 0$, leads to a Euclidean metric).
The chain rule also
provides an additional constraint: namely $r'(0) = 0 \rightarrow
f'(0) = 0$ \cite{Chain}. We may ask that the redshift be a locally increasing
function: $f''(0) \geq 0$. These constitute the minimum requirements.
Beyond this, one could impose a symmetry
on the solutions
of the form $r(l) = r(-l)$ and $f(l) = f(-l)$ which automatically
eliminates all the odd derivatives: $r^{(2n+1)}(0) = f^{(2n+1)}(0) = 0$.
In particular, one can set
$r'''(0) = f'''(0) = 0$. Taken together, these give the boundary
conditions appropriate to a wormhole.
However, one can not choose freely
all the eight boundary conditions independently.
This is easily seen
by writing out the $ll$-equation (6) at the point $l=0$, which yields
an algebraic quartic
equation for the throat radius $r(0)$:
\begin{eqnarray}
-4\left( \frac{f''(0)}{f(0)} \right)^2 [1+\ln(f(0))] r(0)^4
&+& 32\left( \frac{f''(0)r''(0)}{f(0)} \right)[1 + \frac{1}{2}\ln(f(0))]
r(0)^3 \\ \nonumber &+&  [K^{-2}-16 r''(0)^2 \ln(f(0))] r(0)^2
+  16\ln(f(0)) = 0,
\end{eqnarray}
where we have used only that $r'(0) = f'(0) = 0$. Indeed,
it is important to note that this constraining relation is
{\it independent} of the
values assigned to the third derivatives $f'''(0)$ and $r'''(0)$. With some
effort, this quartic equation can be solved for $r(0)$, though the resulting
expressions are not particularly transparent. However, it suffices for our
purposes to consider (8) in certain simplifying but natural cases in order to
get a feeling for the allowed range in $r(0)$. In this way, we have found
that large throat radii can result even for values of
$f(0),f''(0)$ and $r''(0)$
of order unity. As case in point, taking $f(0)=f''(0)=1$, $r''(0)=0$ and
$\xi = \frac{1}{6}$
yields $r(0) \approx 67 l_{P}$ where $l_{P}$ is the Planck length.
Larger (and smaller)
throat radii are also possible depending on the values chosen
for the other boundary
data and the scalar coupling constant $\xi$.

We now turn to our specific calculations of $r(l)$ and $f(l)$
subject to the above
class of boundary conditions.
Employing the Rugge-Kutta method,
we have obtained a good numerical solution of these equations using the
following boundary conditions for $f(l)$ and $r(l)$ at the throat
$l=0$:
\begin{eqnarray}
&&-1\le\ln f(0)<0;\quad f'(0)=0;\quad f''(0)=0;\quad f'''(0)=0;\\
\nonumber
&&r(0)=\sqrt{-16K^2\ln f(0)};\quad r'(0)=0;\quad r''(0)=0;\quad r'''(0)=0
\end{eqnarray}
The graphs which represent a particular numerical solution for $\ln
f(0)=-\frac23$ for various length scales are displayed below.
The redshift function
$f(l)$ is plotted versus proper distance in Figures 1a and 1b; both axes
have been scaled by $K^{-2}$. In Fig. 1b, we see that $f(l)$ is
a positive increasing function. Actually, there is a fine-structure
in the form of Planck-scale ``ripples'' or
spatial oscillations superimposed
on this function; these are not numerical artifacts.
These oscillations are clearly revealed in the
``blow-up'' graph in Fig. 1a. The wormhole's shape function
$r(l)$ is depicted in Figs. 2a and 2b. In Fig. 2a one sees the throat, of
radius $r(0) \approx 0.02\, l_P$,
in the
neighborhood of the origin extending out to about $10K^{-2}$ at which point
the wormhole ``flares'' out, marking the onset of the
superimposed small-scale oscillations.
The graph in Fig. 2b shows the gross features of the shape function.
As remarked below, the oscillations in both $f(l)$ and $r(l)$
are composed of two different modes.
Of course, since both $f$ and $r$ are even functions, their graphs can
be reflected through the origin: $l \rightarrow -l$. The wormhole's
embedding diagram is easily inferred from the Figures 2a and 2b.

It is important to know the asymptotic behavior of $f(l)$ and $r(l)$. To
get at this information, we carried out an asymptotic analysis of the
Einstein equations (5-7) taking into account and guided by
the results of the numerical
investigation. From the numerical calculations, we see that for sufficiently
large $l$, both the redshift and shape functions can be represented as a sum
of two distinct components
$f(l)=F(l)+\phi(l);\quad r(l)=R(l)+\rho(l)$,
where $F(l)$ and $R(l)$ are strictly {\em monotone increasing} and
$\phi(l)$ and $\rho(l)$ are bounded
{\it oscillating} functions.

The relative magnitudes of these components and their derivatives may
be estimated straightforwardly and then used to expand consistently
the coupled Einstein equations. We find that the oscillating modulation
is composed of two modes with frequencies $\omega^2_1 = \frac{1}{16K^2}$
and $\omega^2_2 = \frac{1}{16K^2 (4 + 3 \ln F)}$, respectively, while
in the limit of large $l$, $R(l) \approx l$ and $F(l) \approx
(a\ln l - b)^2$ for $a=5.3$ and $b=25.5$.

From these combined numeric and analytic
calculations we see that
for the chosen set of boundary conditions, $R(l) \rightarrow l$, thus
our self-consistent wormhole 
connects two spatial regions which are asymptotically flat, modulo the
Planck-scale wiggles.
The redshift function however, does not approach a
constant value as $l \rightarrow \infty$,
so the metric as a whole is not asymptotically flat.
We have found additional self-consistent solutions of (1) by taking
different values for the scalar coupling and the boundary data.
In this way, we have found local solutions
which correspond to large throat ($r(0) \approx 200-300\, l_P$)
wormholes with horizons located far from the throat and
wormholes connecting two bounded spatial regions.
A full account of these
calculations will appear in a separate publication.

The work of S.V.S and A.P. was supported in part by the Russian Foundation
of Fundamental Research grant No. 96-02-17366a.

\newpage
\noindent
{{\bf Figure 1a}: The redshift function on small scales}

\noindent
{{\bf Figure 1b}: The redshift function on large scales}

\newpage
\noindent
{{\bf Figure 2a}: The wormhole shape function on small scales}

\noindent
{{\bf Figure 2b}: The wormhole shape function on large scales}


\begin{thebibliography}{99}
\bibitem{Foot1} We have also set the scalar mass and temperature to zero:
$m = \kappa = 0$. In this case, the arbitrary
renormalization scale parameter $\mu$
appearing in $(T^{\nu}_{\mu})_{log}$
can be absorbed into the definition of $f$.
\bibitem{Foot2} This can be checked, for example, by writing the set of
three field equations as an equivalent set of (eight) coupled
first-order equations plus one algebraic equation.
The radial equation reduces to the algebraic transcendental equation in
$f$ and $r$, i.e., it is a constraint.
\end{thebibliography}
\end{document}